# New record in optical gain and room-temperature nanolasers in multiple wavelengths in 2D ErOCl single crystals

**Shipeng Yao[1, 2, 3, †], Hao Sun[1, 3, †, *], Zhang Liang[2], Zhen Wang[2], Lin Gan[1, 3], Jinhua Wu[1, 2, 3], Zhangyu Hou[1, 2, 3], Cun-Zheng Ning[1, 2, 3, *]**

[1]*Department of Electronic Engineering, Tsinghua University, Beijing 100084, China*

[2]*College of Integrated Circuits and Optoelectronic Chips, Shenzhen Technology University, Shenzhen, Guangdong 518118, China*

[3]*International Center for Nano-Optoelectronics, Tsinghua University, Beijing 100084, China*

[†]*These authors contributed equally to this work*

**haosun@mail.tsinghua.edu.cn; ningcunzheng@sztu.edu.cn*

**Abstract:**

Erbium-based materials have long been recognized for their important telecom-band applications, yet their widespread adoption in integrated optoelectronics has been hindered by two fundamental limitations: the difficulty in achieving high erbium density without concentration quenching which leads to small optical gain in doped materials, and the difficulty in fabricating a practical device with single crystal nanowires that demonstrated high optical gain previously[1,2]. Here, we overcome these limitations by synthesizing 2D single crystal ErOCl that has an Er density of $1.75\times10^{22}$ $cm^{-3}$. The high-quality single crystal material significantly reduces the density-related quenching effect that dominates in randomly doped materials with high Er concentration. This results in a record optical gain coefficient over 1500 dB/cm at ~1536 nm band, at least larger by an order of magnitude than the previous gain record in Er materials. Leveraging this exceptional gain medium, we demonstrate room-temperature continuous-wave lasing operation by integrating with a photonic crystal microcavity, achieving a record-low threshold of 7 μW with the most compact size of any Er-based lasers. Furthermore, the unique Stark splitting characteristics of ErOCl provide optical gain in three wavelength bands and lead to lasing in these wavelengths by engineering the cavity. This is the first time that optical gain has been shown in three different wavelength bands in Er materials, together with the smallest size of laser cavity, could have many important applications in on-chip sensing and optical communication.

## Introduction

Advancements in communication, computing, and sensing technologies critically rely on optoelectronic integration. The development of on-chip active devices, particularly nanolasers as essential components of optoelectronic integrated circuits, holds great potential to advance optical interconnects and silicon photonics for next-generation optical systems. These devices not only facilitate high-speed data transmission but also minimize power consumption, seamlessly integrating with existing silicon technologies. However, conventional nanolasers constructed from III-V materials necessitate elaborate and costly heterogeneous integration techniques due to inherent lattice mismatches with silicon[3-5] Furthermore, the temperature-sensitive nature of semiconductor bandgaps constrains device performance and operational versatility. The significant carrier-induced index-gain coupling in semiconductors[6] introduces notable variations in refractive index that degrade emission stability, resulting in frequency fluctuations and increased susceptibility to signal crosstalk[7]. These challenges underscore the urgent need to investigate silicon-compatible materials with enhanced optical gain properties.

Rare earth materials, particularly erbium (Er), have gained prominence as compelling candidates for integrated photonics due to their unique atomic transitions and potential for high optical gain. Erbium provides stable and low-noise emission at 1.5 μm through shielded 4f-4f electronic transitions, which support population inversion and optical amplification while insensitive to environmental perturbations—a critical advantage for minimizing signal crosstalk[8]. However, practical implementation faces a fundamental constraint: achieving efficient and uniform $Er^{3+}$ doping in host matrices. Early approaches employed direct $Er^{3+}$ ion implanting into silicon dioxide ($SiO_2$) microdisk cavities, producing microlasers with diameters of ~ 50 μm[9]. Subsequent advancements in silicon nitride ($Si_3N_4$) and lithium niobate ($LiNbO_3$) microfabrication techniques[10-14] have enabled substantial progress in Er-doped device integration. Liu et al. demonstrated a 3-mm-long Er:$Si_3N_4$ waveguide amplifier with >30 dB gain[15], followed by a photonic-integrated Er laser delivering >10 mW output power and 40-nm-wide wavelength tunability[16]. Wang et al. further achieved room-temperature lasing in a 200-μm-diameter Er: $LiNbO_3$ microdisk with a threshold of ~400 μW[17]. Despite these advances, two critical challenges persist. First, $Er^{3+}$ clustering in randomly doped materials at high doping density reduces optical activity and increases optical losses. Second, the low achievable $Er^{3+}$ concentration ($<10^{20}$ $cm^{-3}$) in conventional platforms necessitate device dimensions exceeding hundreds of micrometers, fundamentally limiting on-chip integration density[1].

Enhancing the gain per unit length of erbium-based materials requires high concentrations, a strategy that has attracted substantial attention[18]. Recent studies reveal that single-crystal erbium chloride silicate (ECS) nanowires exhibit a remarkable net gain of 100 dB/cm at 1.53 μm, marking the highest reported value among erbium-based materials[6]. However, translating these material advancements into functional nanolasers remains a challenge. Primarily, conventional thin-film growth techniques predominantly yield polycrystalline films[19], where grain boundaries induce severe non-radiative recombination, degrading optical efficiency. Furthermore, the relatively low refractive indices of erbium compounds (~1.7) demand precise integration with high-index contrast cavity structures to amplify light-matter interactions. While chemical vapor deposition (CVD)-grown single-crystal erbium compound nanowires exhibit superior crystallinity, their inherent morphological variability complicates post-processing steps required for optimized microcavity coupling[20]. These limitations underscore the critical need to exploit intrinsic optical properties and structural varieties in erbium compounds while developing scalable integration protocols for on-chip cavity architectures.

Two-dimensional (2D) materials offer unprecedented opportunities for integrated photonic systems due to their atomic-scale confinement and tailorable light-matter interactions[21-23]. Among rare-earth compounds, lanthanide oxyhalides (LnOX, X=Cl, Br, I)[24,25] — particularly those incorporating heavier lanthanides (Ho, Er, Tm, Yb, Lu) — exhibit van der Waals layered structures[25-27] that facilitate direct integration with heterogeneous substrates. Recent advances include the large-scale growth of ErOCl[28] , where the 2D structure ensures uniform $Er^{3+}$ distribution at high concentrations ($1.75\times10^{22}$ $cm^{-3}$) while suppressing concentration quenching through symmetry-protected 4f transitions. Despite the successful synthesis of ErOCl and other lanthanide oxyhalides, their practical application demands with rigorous quantification of intrinsic optical gain—a critical parameter previously underexplored.

In this study, single-crystalline ErOCl nanosheets were synthesized on Silicon substrates utilizing sodium chloride (NaCl) as a catalyst. We employ differential reflectance spectroscopy (DRS) to systematically resolve the gain dynamics of ErOCl. The DRS measurements reveal a peak gain coefficient of over 1500 dB/cm at ~1536 nm band, surpassing conventional erbium compounds by two orders of magnitude. This unprecedented gain performance originates from simple and symmetrical 2D lattice structure with less non radiative recombination channels, in which Er ions maintain a nearly intrinsic energy level structure. Leveraging this fundamental advance, we integrated these high-gain ErOCl crystals with silicon photonic crystal microcavities, achieving room-temperature continuous-wave lasing at 1.5 μm with a record-low threshold of 7 μW. The intrinsic Stark-split gain spectrum in ErOCl further enables wavelength-controllable lasing from 1500 nm to 1565 nm on a single silicon-on-insulator (SOI) chip through cavity-mode engineering, providing the potential to fulfill the critical need for on-chip wavelength-division multiplexing (WDM). These achievements position 2D rare-earth materials as the first viable gain medium class combining atomic transition stability with photonic circuit scalability – a critical leap toward on-chip optical interconnects.

## Results

### Growth of Single-Crystal ErOCl on Silicon

Single-crystalline ErOCl nanosheets were synthesized on silicon substrates at 900 °C via CVD. The introduction of NaCl as a growth promoter significantly enhanced nucleation kinetics and reaction efficiency, yielding highly regular hexagonal nanosheets with well-defined edges (Fig. 1a). Our experiments revealed that trace amounts of NaCl (approximately one-ninth of the total precursor mass) initial nucleation above 800 °C but rapidly evaporate post-nucleation due to thermal instability under elevated temperatures (see SI, Fig. S3k). Parametric studies varying NaCl concentrations and growth durations (See SI, Figure S3) provided mechanistic insights into the vapor-phase synthesis: NaCl acts as a catalyst that lowers nucleation energy barriers while avoiding incorporation into the final crystalline lattice due to its thermal volatility. Phase purity and crystallographic orientation of the as-grown nanosheets were confirmed by X-ray diffraction (XRD), with all diffraction peaks matching the standard ErOCl reference (JCPDS No. 49-1800) (Fig. 1b). The hexagonal morphology of ErOCl nanosheets is evident in the scanning electron microscopy (SEM) image shown in the inset of Fig. 1b, displaying edge lengths of 5 μm.

To probe the atomic-scale crystallinity and stacking configuration, high-resolution transmission electron microscopy (HRTEM) was systematically employed. Longitudinal HRTEM imaging (Fig. 1c) resolves well-ordered atomic arrangements along the [110] crystal direction, with a measured interplanar spacing of 0.19 nm. Transverse cross-sectional analysis (Fig. 1e) unveils a periodic van der Waals layered architecture along the [001] direction, exhibiting monolayer thicknesses of 0.88 nm, consistent with theoretical interlayer distances, thereby validating that the interlayer connectivity is maintained by van der Waals bonds. The corresponding selected area diffraction pattern (SADP) acquired from distinct zone axes in Fig. 1d and Fig. 1f further validates the single-crystal nature and hexagonal close-packed (HCP) symmetry. Complementary atomic force microscopy (AFM) profiling (see SI, Fig. S1c) quantified the stepwise growth morphology, revealing monolayer terraces with thickness of ~ 0.9 nm, consistent with HRTEM observations.

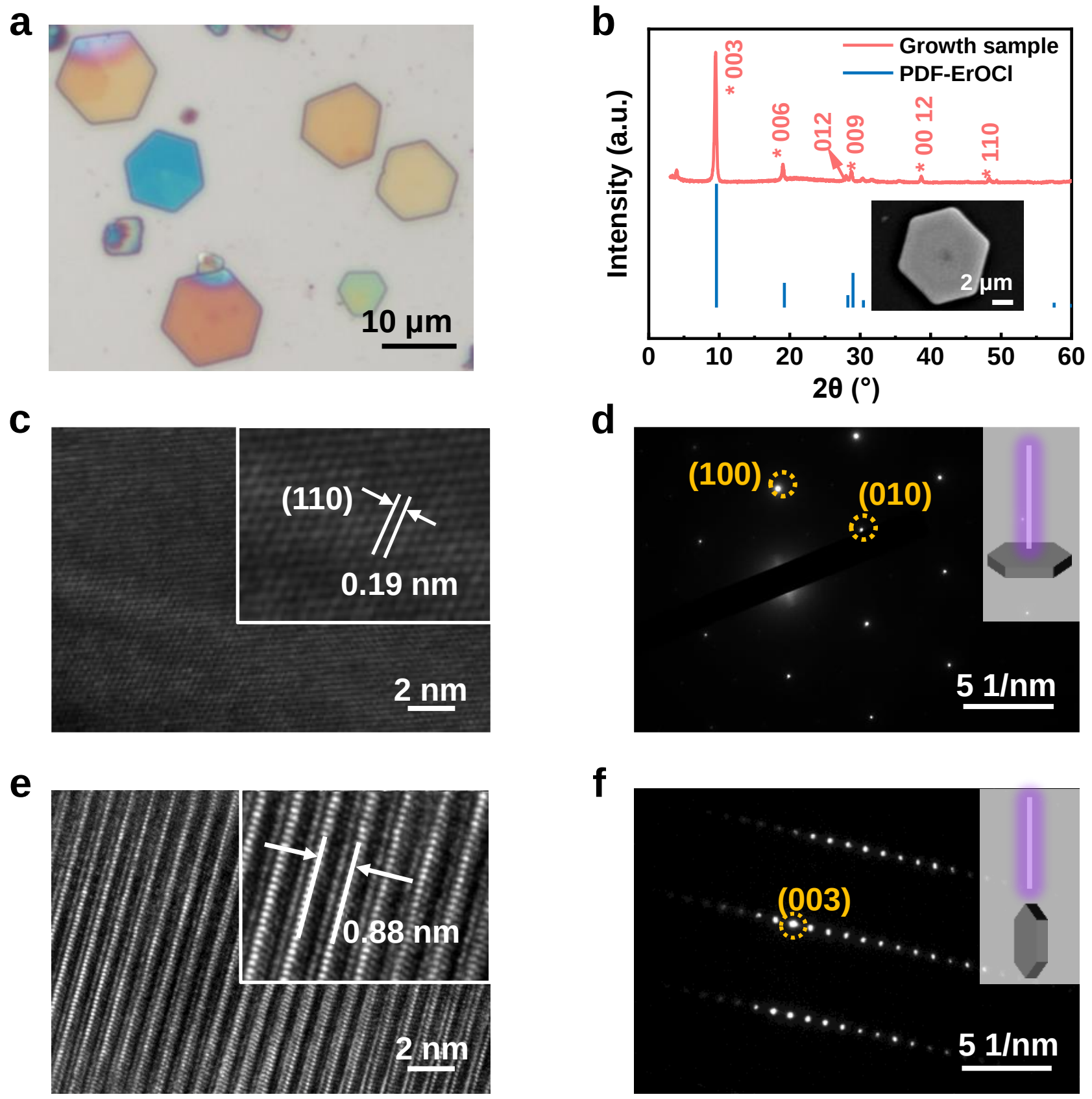

Figure 1. Phase and Crystallographic Characterization of ErOCl nanosheets. (a) Optical microscopy image of as-grown hexagonal ErOCl nanosheets on a silicon substrate. (b) XRD spectrum of the synthesized sample (red) compared to the reference ErOCl pattern (JCPDS No. 49-1800). Inset: SEM image showing the regular hexagonal morphology of a representative nanosheets. (c) HRTEM image of the longitudinal cross-section, revealing atomic arrangements with 0.19 nm lattice spacing. (d) Corresponding selected-area diffraction pattern confirming single-crystal structure. (e) Transverse cross-sectional HRTEM image showing the layered van der Waals structure with 0.88 nm interlayer spacing. (f) Corresponding diffraction pattern of the zone in (e).

**Optical Gain Characterization of ErOCl**

The intrinsic optical properties of ErOCl were systematically investigated using a customized micro-photoluminescence (μ-PL) system optimized for the 1.5 μm communication band. Figure 2a depicts the distinct transition process between sub-energy levels corresponding to various emission bands spanning 1500-1560 nm under 980 nm laser excitation. These radiative transitions facilitate synchronized emission from $Er^{3+}$ ions ensembles, generating intense spectral outputs with exceptional line concentration, in stark contrast to the inhomogeneous broadening profiles typically observed in disordered erbium-doped materials. Room temperature PL measurements reveal well-resolved Stark-level splitting and nanometer scale emission linewidths, as show in Fig. 2b, demonstrating exceptional spectral purity with a main peak full width at half maximum (FWHM) below 3 nm. This narrow-line emission originates from the uniform periodic potential of the crystalline lattice, which maintains exceptional site symmetry for $Er^{3+}$ ions. Temperature-dependent spectral evolution under 980 nm excitation (Supporting Information Fig. S9a) demonstrates remarkable thermal stability, maintaining narrow emission profiles from cryogenic (8 K) to near-ambient (280 K) conditions. Spectral analysis reveals temperature-dependent broadening of the 1537 nm spontaneous emission band, accompanied by emerging high-intensity peaks at 1504 nm and 1562 nm with increasing thermal energy.

The realization of population inversion between the $^4I_{13/2}$ and $^4I_{15/2}$ Stark manifold transitions (governing 1.5 μm emission) necessitated precise quantification of the 980 nm absorption cross-section. We adopted a differential reflectance methodology employing ErOCl nanosheets transferred onto gold-coated silicon substrates (100 nm Au coating), as illustrated in SI, Fig. S5a. Broadband spectral analysis revealed comparable absorptions magnitudes at 980 nm and 1.5 μm (Fig. 2c). Through transfer matrix modeling of multilayer optical interference[29,30] (see SI, Section 2.2), we determined an absorption cross-section of ~ $6.28\times10^{-21}$ cm² at 980 nm, representing a 2-3 times enhancement over conventional erbium-doped systems. Complementary characterization of the 1.5 μm absorption through transmittance spectroscopy and differential reflectance spectroscopy confirmed the exceptional absorption nature of the crystalline lattice. Remarkably, when transferred onto the cross-section of a single-mode optical fiber (see SI, Fig. S4), the ErOCl nanosheets exhibited pronounced absorption maxima at 1504 nm and 1536 nm, achieving absorption coefficients of 1934 dB/cm at 1536 nm (see SI, Fig. S4d). This large 1.5 μm absorption resonance, coupled with the enhanced 980 nm excitation efficiency, establishes ErOCl as a superior gain medium for C-band optical amplification.

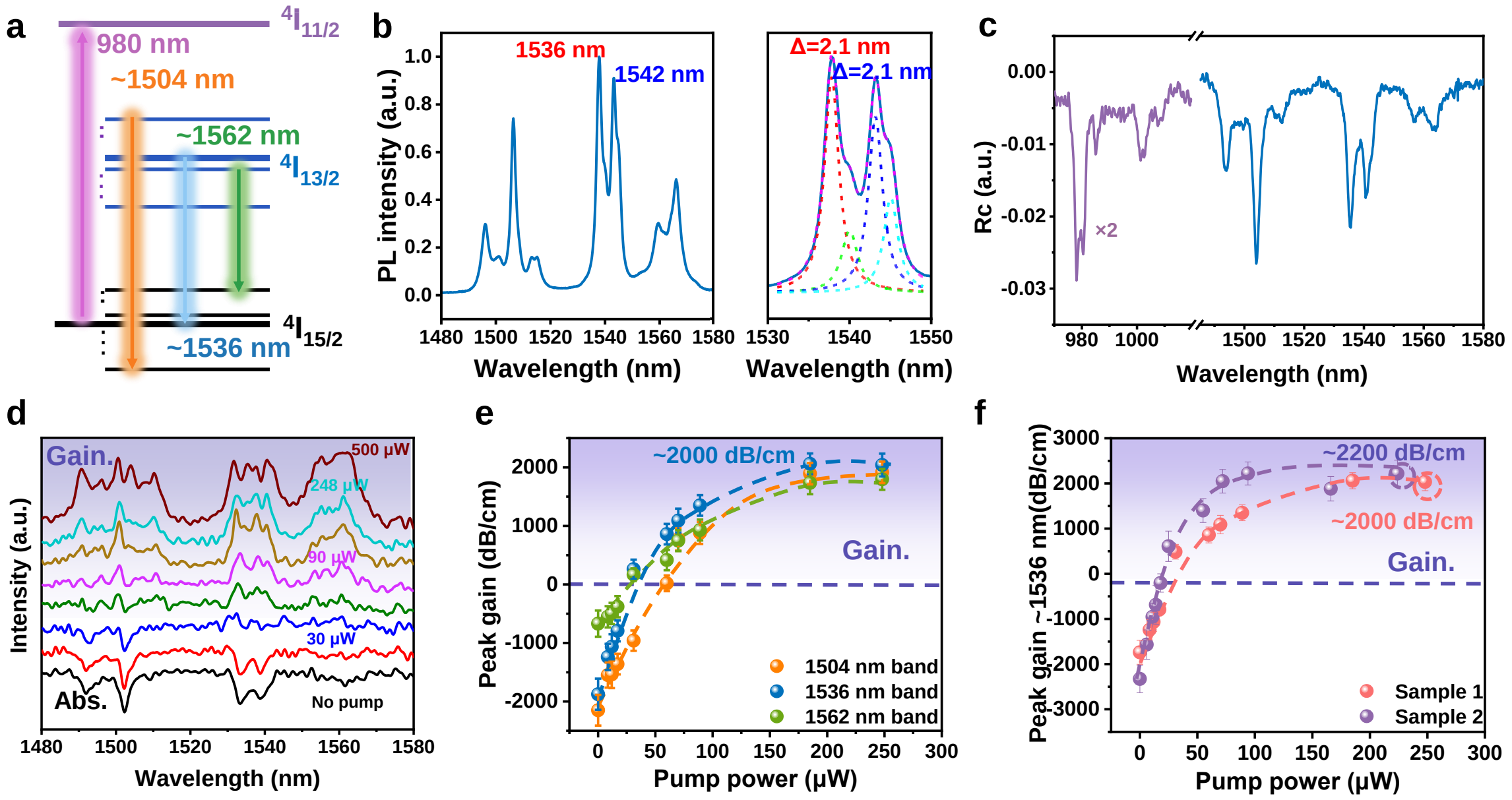

Figure 2. **Optical properties and gain characteristics.** (a) Energy-level diagram illustrating Stark-split sublevels and corresponding radiative transitions. (b) The 1.5 μm-band PL spectrum under 980 nm excitation at room-temperature. Inset: PL spectrum from 1530-1550 nm showing FWHM of the main peak is sub-3nm. (c) Differential reflectance spectrum under broadband illumination, comparing absorption at 980 nm and 1.5 μm. (d) Power-denpendent differential reflectance spectra under 980 nm excitation. (e) Fitted gain coefficient derived from power-dependent measurements, demonstrating that the gain gradually saturates when it exceeds 1500 dB/cm. (f) Gain growth curves of two samples in the 1530 nm wavelength band.

The optical amplification characteristics of ErOCl was investigated through pump-probe spectroscopy employing 980 nm excitation and broadband detection (see SI, Fig. S5b). We developed a differential reflectance metric Rc to quantify net optical gain[21]:

$Rc = \frac{R(p,b,s) - R(p,0,s) - R(0,b,0)}{R(0,b,0)}$, where R(p,b,s) represents composite reflectance under concurrent pump-probe illumination ("p" for "pump" , "b" for "broadband probe", "s" for "sample"); R(p,0,s) represents amplified spontaneous emission under pump-only illumination; and R(0,b,0) corresponds to the reflectance of gold substrate reference. Figure 2d demonstrates pump-dependent spectral evolution, showing characteristic absorption minima at 1530 nm and 1560 nm progressively converting to gain maxima. The detailed data processing procedure is provided in Section 2.3 of the Supplementary Information (SI) and illustrated in Fig. S7. This provides direct evidence of low-threshold population inversion ($P_{th}$< 30 μW) between Stark-split $^4I_{13/2} \rightarrow$ $^4I_{15/2}$ sublevels. The observed inversion hierarchy directly correlates with crystal field splitting energetics: higher-energy transitions (1500 nm band) require increased pumping densities due to larger inter-sublevel energy gaps. Beyond the 1500 nm gain threshold, continued pump increase drives the system through sequential operational phases - initial linear gain growth transitions into a homogeneous broadening regime at elevated powers (P> 90 μW), where thermal redistribution of populations across Stark sublevels generates broad gain bands spanning 1500-1560 nm. Further increasing the pump power leads to broadening of these gain peaks, a signature of saturation effects where higher energy states become increasingly populated. At even higher pump powers (> 185 μW), nonlinear up-conversion processes occur and competes with the down-conversion process, leading to a pronounced reduction in gain within the 1530 nm band. This cascaded spectral evolution manifests as characteristic lineshape transformation from discrete peaks to continuous bands, governed by accelerated inter-sublevel relaxation dynamics under high excitation densities.

Systematic analysis of pump-dependent gain spectra in Fig. 2e reveals distinct wavelength-selective amplification behavior, where shorter wavelength transitions (~1500 nm band) exhibit higher gain thresholds and reduced peak gain intensities compared to their longer wavelength counterparts. This asymmetric gain distribution demonstrates an inverse relationship between operational wavelength and pumping efficiency. Figure 2f displays the peak gain growth curves of two distinct samples in the 1536 nm emission band. Both samples demonstrate saturation peak gain exceeding 2000 dB/cm, representing nearly two orders-of-magnitude improvement over conventional erbium-doped gain media. The reasonable variations observed are attributed to differences in the sample's intrinsic crystalline quality and luminescence properties, as discussed further in SI, Section 2.4. The 980 nm-pumped differential reflectance spectra quantitatively confirm ErOCl's exceptional amplification capacity spanning 1500-1560 nm window. This broadband amplification capability, combined with its silicon-compatible crystalline structure, positions ErOCl as an unprecedented platform for developing energy-efficient nanophotonic devices.

**Nanolaser Characterization and Performance**

Building upon the exceptional gain coefficients demonstrated in broadband spectroscopy, we proceeded to validate these optical properties through nanolaser implementation. The realization of silicon-based nanolasers necessitates high-quality microcavities capable of well-coupled integration with active gain media[21,31-33]. Leveraging the anisotropic structural properties of 2D ErOCl nanosheets, we engineered L3-type photonic crystal (PC) microcavities (theoretical quality factor $Q>10^4$) optimized for dual functionality: subwavelength light confinement and deterministic material-cavity coupling. Finite-difference time-domain (FDTD) simulations established critical design parameters---material thickness falls within 30 to 60 nm, with a refractive index of 1.7 at ~1.5 μm (see SI, Figure S10). Through van der Waals assembly techniques, we achieved precision alignment of ErOCl nanosheets onto the cavity mode spot. The deterministic integration process, detailed in SI Section S3.2, preserves both the crystalline integrity of ErOCl and the photonic band structure of PC cavities - a critical prerequisite for observing low-threshold lasing behavior.

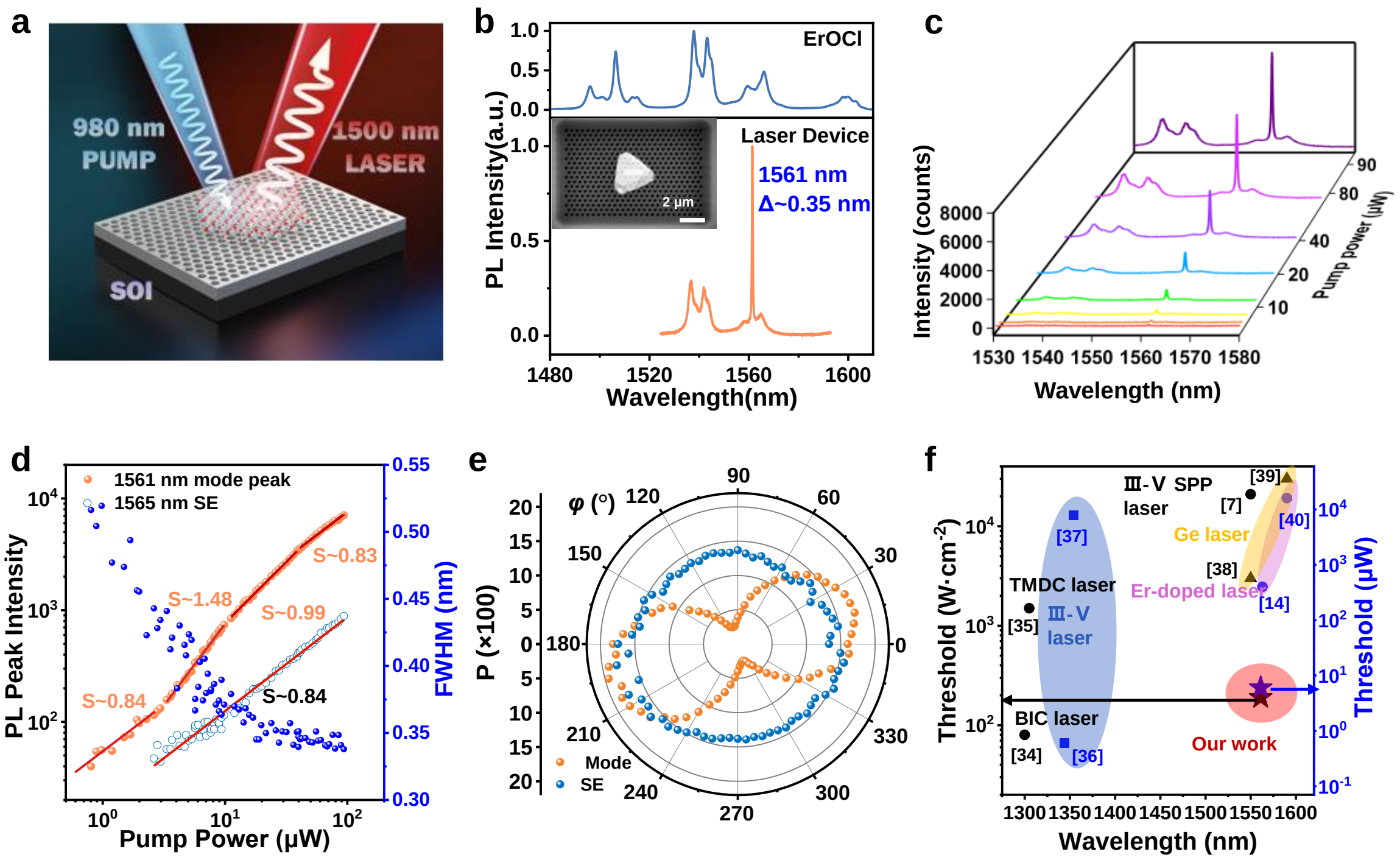

Figure 3 Nanolaser device performance. (a) Schematic of the photonic crystal (L3 cavity)- integrated ErOCl nanolaser. (b) Comparative PL spectra: ErOCl nanosheet emission (top) vs. cavity-coupled lasing output (bottom). (Inset:SEM image showing precise alignment of an ErOCl nanosheet within the cavity region.) (c) Lasing spectra at different pump powers. (d) L-L curve showing nonlinear threshold behavior (black dots: lasing mode; open circles: spontaneous emission). Blue dots track linewidth narrowing from 0.52 nm to 0.34 nm. (e) Polarization-resolved emission: linear polarization of lasing mode (orange dots) vs. unpolarized spontaneous emission (blue dots). (f) Benchmark comparison of lasing thresholds with state-of-the-art 1.3-1.5 μm nanolasers.

Building upon the engineered photonic crystal cavity architecture, we characterized lasing performance through micro-PL spectroscopy under ambient conditions. The experimental configuration (illustrated in Fig. 3a) employed a 980 nm continuous-wave laser as the excitation source. Near-infrared emissions were collected through a 100× objective and analyzed using a high-resolution detection system (0.06 nm), enabling precise discrimination between spontaneous and stimulated emission regimes. Spectroscopic analysis indicates essential distinctions between intrinsic material emission and emission resulting from material-cavity coupling. While uncoupled ErOCl exhibits characteristic 1537 and 1542 nm spontaneous emission peaks (Fig. 3b, upper trace), the cavity-coupled device demonstrates a prominent lasing mode at 1561 nm (lower trace). SEM imaging (Inset of Fig. 3b) confirms the deterministic coupling between a sub-50 nm-thick ErOCl nanosheet and the L3 cavity's electric field maximum.

Figure 3c presents pump-dependent evolution of PL spectra, revealing the characteristic competition between spontaneous emission (SE) and stimulated mode formation. Quantitative analysis of pump-dependent intensity evolution reveals three distinct operational phases (Fig. 3d). Below the lasing threshold (<7 μW, ~220 W/cm²), the system operates in a subthreshold regime demonstrating near-linear intensity scaling with a slope parameter S=0.84. As pump power increases through the transition region (7-20 μW), we observe pronounced nonlinear behavior manifesting as superlinear amplification (S=1.48) – a characteristic signature of stimulated emission onset. The system enters stable lasing operation between 20-40 μW (S=0.99), maintaining linear intensity progression until thermal effects and excited-state absorption mechanisms dominate above 40 μW, inducing gain saturation (S=0.83) in the high-power regime. In contrast, the spontaneous emission near 1565 nm exhibits a sublinear increase with a slope of 0.84 across the entire pump power range. Spectral linewidth analysis offers critical evidence for the onset of lasing, with the dominant mode narrowing from 0.52 nm to 0.34 nm as the pump power surpasses the threshold (see SI Section S4 for detailed fitting results).

Polarization-resolved spectroscopy (Fig. 3e) further demonstrate the evolution from isotropic spontaneous emission (circularly polarized) to linearly polarized stimulated emission, governed by the cavity mode selection rules. Complementary measurements in SI Sections 5 quantify the emergence of spatial coherence through real-space mode profile evolution, while Section 6 reveals accelerated recombination dynamics with emission lifetime reduction from 1230 μs (below-threshold) to <50 μs (above-threshold) - a 20-fold decrease consistent with stimulated emission dominance. The concurrent observation of nonlinear L-L threshold behavior, spectral linewidth narrowing, polarization anisotropy development, and emission lifetime quenching, provides unambiguous signatures of coherent light generation in this van der Waals integrated system.

Figure 3f presents a comparative analysis of our device against state-of-the-art silicon-integrated nanolasers by comparing threshold densities[34-40], including III-V lasers[25,26], Ge lasers[27,28], surface plasmon polariton (SPP) lasers[5], and Er-based lasers[12,29] (see SI, Table 2 for detailed information). Our ErOCl-based device establishes a low threshold of 7 μW (220 W/cm²) at 1.5 μm, representing more than 100× improvement over conventional erbium gain systems. This unprecedented efficiency stems from two synergistic advances: the intrinsically large gain coefficients (>1500 dB/cm) of the 2D ErOCl crystals, and strong light-matter interaction caused by small mode volume in L3 microcavities. The detailed rate equation modeling and β-factor extraction can be seen in SI, Section 7. The two-orders-of-magnitude threshold reduction, compared to erbium-doped dielectrics[17,41,42], redefines the scalability limits for integrating high-density laser arrays in silicon photonics, addressing the critical power bottleneck in optical interconnects.

**Wavelength-controlled Single-mode Lasing through Cavity Design**

In light of growing demand for high-capacity optical information processing and storage, the integration of different wavelength laser sources onto a single chip can significantly reduce associated packaging costs. The effective collaboration among multiple devices relies on stable emission from the gain medium, single-mode properties of each laser, and compatible fabrication processes. Our microcavity device employing ErOCl as the gain material embodies these advantages and maintains exceptional stable and narrow linewidth emission (<0.4 nm), which is maintained for over 80 days under room temperature conditions. (SI, Fig. S19a). Furthermore, ErOCl demonstrates broad splitting band, stable spectral characteristics, positioning it to facilitate wide-range, wavelength-controllable lasing through cavity design. We designed and fabricated an array of microcavities with mode variations across the 1.5 μm range by adjusting the radius and periodicity of holes in the L3 microcavity. The precise transfer of thin ErOCl layers of appropriate size and thickness to distinct microcavities enabled the formation of wavelength-controllable nanolasers (Fig. 4a).

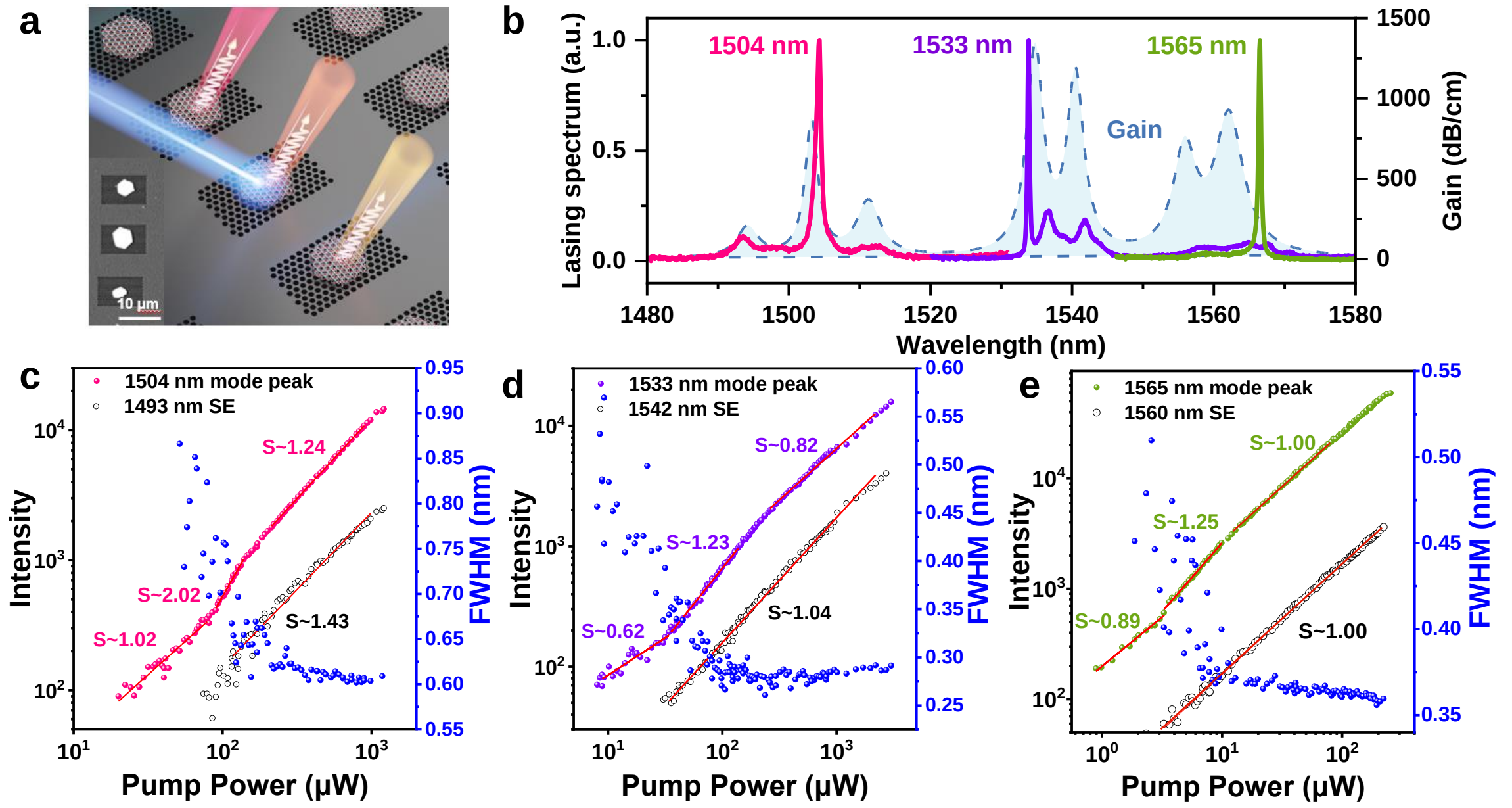

Figure 4 Wavelength-controllable nanolaser array. (a) Multi-Wavelength Chip Schematic. Inset: optical micrograph of 3 ErOCl-integrated nanolasers on the same chip. (b) Emission spectra from devices with lasing wavelengths spanning 1504-1565 nm and the gain spectrum of ErOCl before saturation (the blue dashed line); (c)-(e) L-L curves and corresponding linewidth evolution for devices operating at (c) 1504 nm (highest threshold), (d) 1531 nm (optimal efficiency), (e) 1565 nm (lowest threshold).

Systematic characterization of the laser array revealed coherent emission spanning 1500–1565 nm (Fig. 4b). The blue shaded region represents the gain spectrum of ErOCl just before the onset of significant saturation, clearly showing broadband gain across three distinct bands. The threshold power at 1504 nm (~ 120 μW) and 1533 nm (~ 70 μW) is higher than that of 1560 nm (<10 μW) by the light–light (L–L) curves and linewidth evolution in Figs. 4c–e. This spectral dependence of lasing thresholds directly correlates with the energy hierarchy of Stark-split sublevels in ErOCl's $^4I_{13/2} \rightarrow {}^4I_{15/2}$ transition. Higher-energy transitions at shorter wavelengths (1504 nm) require larger pumping intensity to achieve population inversion compared to longer-wavelength counterparts (1565 nm), as theoretically predicted by the gain spectrum in Fig. 2e. The observed linewidth narrowing coupled with superlinear L–L curves across all wavelengths, provides unambiguous evidence of stimulated emission dominance. These findings validate the efficacy of our approach in achieving wavelength-controllable lasing capabilities across targeted gain wavelength ranges.

## Conclusions

We achieved the growth of single-crystalline 2D ErOCl nanosheets on silicon substrates through a NaCl-assisted CVD process. The incorporation of NaCl effectively enhances growth rates and promotes both the nucleation and synthesis reactions, facilitating the formation of regular hexagonal nanosheets. Notably, the material exhibits a saturation gain coefficient exceeding 1500 dB/cm, representing more than one order-of-magnitude improvement over the previous gain record we achieved in Er compound nanowires[6]. This new gain record, combined with the van der Waals-layered architecture, offers a readily integrable approach with a cavity for achieving strong light-matter interactions. Leveraging these advantages, we integrated ErOCl nanosheet with a photonic crystal microcavity to demonstrate room-temperature continuous-wave lasing at 1.5 μm. The low threshold of 7 μW is two orders of magnitude lower than conventional erbium-doped systems, confirming the exceptional gain properties of ErOCl. To the best of our knowledge, the presented laser represents the most compact rare-earth-based laser yet developed. The exploration of 2D rare-earth materials and the successful implementation of silicon-based nanolasers not only provide viable solutions to the challenges posed by on-chip interconnects, but also pave the way for the realization of efficient integrated photonic devices.

In addition, we would like to highlight the capacity of our nanolaser for wavelength-controllable lasing within three bands between 1500 nm and 1565 nm, due to the specific crystal field effects in ErOCl. Such multi-wavelength nanolasers has the potential to overcome limitations inherent to traditional communication systems for novel on-chip apllications. Stable and separated gain bands are expected to find applications in on-chip amplifiers in photonic circuits, further enhancing compatibility with diverse communication standards and opening up new possibilities for integrating nanolasers in multispectral photonics applications. This materials breakthrough combined with CMOS-compatible microcavity fabrication process, establishes a manufacturable pathway toward large-scale photonic circuits containing >20000 lasers/mm². Such integrations potentially yield improved device performance, enhanced efficiencies, and novel functionalities.

## Material and methods

### CVD Growth of 2D ErOCl Single Crystals

The synthesis of two-dimensional erbium oxychloride (2D ErOCl) was achieved via chemical vapor deposition (CVD) at 900 °C. A mixed precursor containing anhydrous $ErCl_3$ particles (99.9%, Alfa Aesar) and NaCl power (99.5%, Alfa Aesar) in an 8:1 weight ratio was positioned in the high-temperature zone of the CVD system. During growth, silicon (111) substrates were positioned face-down upstream from the precursor source to establish optimal vapor-phase transport conditions. The thermal process commenced with a controlled temperature ramp (20 °C/min) to 900 °C under 95% Ar/5% $H_2$ carrier gas flow, followed by a 1-hour isothermal growth phase. Systematic parameter studies revealed a critical minimum growth duration of 20 minutes for single-crystalline domain formation.

### Structural characterization

Crystallographic analysis was conducted via grazing-incidence X-ray diffraction (GIXRD, Rigaku D/max-2500/PC) using Cu Kα radiation at a fixed incident angle of 3° to suppress substrate interference from the Si (111) growth substrate. Phase identification was achieved through comparison against reference patterns JCPDS No. 49-1800. Complementary atomic-scale investigations employed field-emission transmission electron microscopy (FE-TEM, JEOL JEM-2100F) operated at 200 kV with point resolution <0.1 nm. Mechanically exfoliated ErOCl flakes were transferred onto carbon-coated Cu grids via deterministic fiber-assisted transferring technique.

**Optical spectroscopy**

Photoluminescence (PL) spectroscopy and differential reflectance spectroscopy (DRS) measurements were performed using a microscopy system with backscattering geometry, incorporating a 100× near-infrared optimized objective lens (NA=0.7). Spectral dispersion was achieved via Princeton Instruments SP-2560 spectrometer equipped with liquid nitrogen-cooled detectors: InGaAs or Si CCD (PYLON-400BRX) for near-infrared or visible wavelengths. A InGaAs array detector (NIRvana ST640) was used for spatially resolved PL mapping. The excitation source comprised a thermally stabilized 980 nm diode laser controlled by a Thorlabs ITC4001 driver, providing both continuous-wave and pulsed operation modes.

For time-resolved PL dynamics analysis, a synchronized measurement setup was implemented combining a 980 nm pulsed laser (200 μs pulse width, 600 Hz repetition rate) with a PicoQuant Hydraharp 400 time-correlated single-photon counting (TCSPC) system. Emitted photons were collected through a multimode optical fiber after passing through a 1000 nm long-pass dichroic filter to suppress residual excitation light. Detection was alternately performed using two optimized sensors: a Micro Photon Devices InGaAs avalanche photodiode (PDM series), and a superconducting nanowire single-photon detector exhibiting 80% quantum efficiency at 1550 nm.

**Numerical Modeling and Device Fabrication**

Optical mode analysis of the L3 photonic crystal cavity was conducted through three-dimensional finite-difference time-domain (FDTD) simulations. Systematic parameter sweeps evaluated the quality factor (Q) dependence on ErOCl active layer thickness, employing perfectly matched layer (PML) boundary conditions and broadband dipole sources. Device realization commenced with electron-beam lithography (EBL) patterning on 220 nm silicon-on-insulator (SOI) substrates using 100 kV acceleration voltage and 200 pA beam current for critical dimension control. The defined nanostructures were transferred via inductively coupled plasma reactive ion etching (ICP-RIE). Subsequent buffered oxide etch (BOE) process for 8–11 minutes selectively removed the underlying $SiO_2$ layer, followed by critical point drying to prevent stiction-induced structural collapse.

## Acknowledgment

The authors acknowledge financial support from the National Key R&D Program of China (Grant Nos. 2021YFA1400700, 2021YFA1401200, and 2023YFF1500802); National Natural Science Foundation of China (Grant Nos. 62433006 and 61975252); Pingshan Innovation Platform Project of Shenzhen Hi-tech Zone Development Special Plan in 2022 (29853M-KCJ-2023-002-01); Universities Engineering Technology Center of Guangdong (2023GCZX005); Natural Science Foundation of Top Talent at SZTU (GDRC202301); Engineering Research Center of Guangdong for Compound Semiconductor Devices and Chips. Key Programs Development Project of Guangdong (2022ZDJS111); and Frontier Science Center for Quantum Information.

## Author contribution

H.S. and S.Y. conceived and designed the experiments. S.Y. performed the material synthesis and device fabrication. H.S. and S.Y. optimized the optical system and conducted the gain and laser characterization measurements. Z.L. and S.Y. built the CVD system. Z. L guided Y.S. in the early stages of material growth experiment. Z.W. performed absorption measurements and guided the gain spectra fitting. L.G., S.Y., and J.

W. designed and optimized the photonic crystal microcavity. J. W. and Z.H. contributed to the discussions regarding material characterization and the rate equation modeling. S.Y. processed the experimental data under the guidance of H.S. S.Y., H.S., and C.Z.N. performed the data analysis and wrote the manuscript. H. S. and C.Z.N. co-supervised the overall project.

## Additional information

Supplementary Information is available for this paper. Reprints and permissions information is available at www.nature.com/reprints. Correspondence and requests for materials should be addressed to H.S. and C.Z.N.

## Competing financial interests

The authors declare no competing financial interests.